\documentclass[submission,copyright,creativecommons]{eptcs}
\providecommand{\event}{MBT 2013} 
\usepackage{breakurl}             
\usepackage{amsfonts}
\usepackage{amssymb}
\usepackage{amscd}
\usepackage{amsmath}
\usepackage{euscript}
\usepackage{graphicx}
\usepackage{algorithm}
\usepackage{algpseudocode}
\usepackage{algorithmicx}

\usepackage{float}
\usepackage{color}

\floatname{algorithm}{Action}

\newtheorem{definition}{Definition}{\bfseries}{\bfseries}
\newtheorem{example}{Example}{\bfseries}{\bfseries}
\newtheorem{statement}{Statement}{\bfseries}{\bfseries}

\graphicspath{{./figures/}}

\title{Runtime Verification Based on Executable Models:\\
On-the-Fly Matching of Timed Traces}
\author{Mikhail Chupilko \quad\qquad Alexander Kamkin
\institute{Institute for System Programming of the Russian Academy of Sciences (ISPRAS)\\
109004, Russia, Moscow, Alexander Solzhenitsyn st., 25.}
\email{\{chupilko,kamkin\}@ispras.ru}
}
\def\titlerunning{Runtime Verification Based on Executable Models: On-the-Fly Matching of Timed Traces}
\def\authorrunning{M. Chupilko \& A. Kamkin}
\begin{document}
\maketitle
\begin{abstract}
Runtime verification is checking whether a system execution satisfies or violates a given correctness property.
A procedure that automatically, and typically on the fly, verifies conformance of the system's behavior to the specified property is called a monitor.
Nowadays, a variety of formalisms are used to express properties on observed behavior of computer systems, and
a lot of methods have been proposed to construct monitors.
However, it is a frequent situation when advanced formalisms and methods are not needed, because an executable model of the system is available.
The original purpose and structure of the model are out of importance;
rather what is required is that the system and its model have similar sets of interfaces.
In this case, monitoring is carried out as follows.
Two ``black boxes'', the system and its reference model, are executed in parallel and stimulated with the same input sequences;
the monitor dynamically captures their output traces and tries to match them.
The main problem is that a model is usually more abstract than the real system, both in terms of functionality and timing.
Therefore, trace-to-trace matching is not straightforward and allows the system to produce events in different order or even miss some of them.
The paper studies on-the-fly conformance relations for timed systems
(i.e., systems whose inputs and outputs are distributed along the time axis).
It also suggests a practice-oriented methodology for creating and configuring monitors for timed systems based on executable models.
The methodology has been successfully applied to a number of industrial projects of simulation-based hardware verification.
\end{abstract}

\section{Introduction}

{\it Verification} has long been recognized as one of the integral parts of software and hardware design processes~\cite{sw-verification,hw-verification}.
Generally, it is an activity intended to check whether a system or its part meets a {\it specification} (set of functional and timing requirements).
Verification techniques can be divided into two main groups, namely {\it formal verification} and {\it testing}
(also known as {\it simulation-based verification} in the hardware engineering domain)~\cite{lam}.
Formal methods are aimed at rigorous proving or disproving the correctness of a formal model of a system with respect to a formal specification.
Such approaches exhaustively examine all possible executions of a given system --
either explicitly (by enumerating all reachable states) or
implicitly (by using symbolic techniques).  
In contrast, testing deals with a finite number of executions and estimates the system's behavior in a finite number of situations (so-called {\it test situations}).
{\it Runtime verification} is a common point of both.
Like testing, it works with concrete executions of a system, but does it in a formal way.

In runtime verification, a correctness property is typically expressed in a formal language,
which makes it possible to automatically translate the property into a {\it monitor}.
Such a monitor is then used to check a system execution with respect to that property~\cite{bauer-2011}.
The idea is similar to {\it specification-based testing},
where a formal specification serves as a basis for generating a {\it test oracle},
which, like monitor, determines whether an observed behavior is consistent with the specification~\cite{spec-testing-2009,contract-specs-2007}.
But, as opposed to testing, it is not a scope of runtime verification to construct test sequences and apply them to the system under test.
The task is to passively observe inputs and outputs of the system and to check their conformance~--
that is why it is also called {\it passive testing}~\cite{passive-testing-2012}.
Formally, when $\EuScript{L}(\varphi)$ denotes the set of valid system executions given by property $\varphi$,
runtime verification is aimed at checking whether a concrete execution $w$ is an element
of $\EuScript{L}(\varphi)$. In this sense, runtime verification deals with the {\it word
problem}, i.e., identifying whether a given word is included in some language~\cite{bauer-2011}.

Correctness properties in runtime verification may be expressed using a variety of different formalisms, including
{\it extended regular expressions}~\cite{sen-2003},
{\it contract specifications}~\cite{contract-specs-2007} and
{\it rule-based approaches}~\cite{barringer-2007}.
{\it Temporal logic}, which is well-known from {\it model checking}~\cite{model-checking},
is also very popular in runtime verification, especially variants of {\it linear temporal logic},
such as LTL and TLTL (a natural counterpart of LTL in the timed setting)~\cite{bauer-2011}.
There are also a lot of methods for generating effective monitors (or test oracles) from formal specifications.
However, sophisticated formalisms and methods are not always suitable for industrial practice.
For example, many hardware design companies use {\it executable software models} for design space exploration and architecture modeling;
it is quite natural to {\it reuse} those models for verification and monitoring.
High reusability within a project is important to complete verification within the timeline~\cite{monitor-based}.
Moreover, reusable models ensure conceptual integrity of the project and accelerate the knowledge interchange.

Runtime verification based on executable models is carried out in the following way.
A reference model is {\it co-executed} with the target system and applied with the same inputs as the system under verification.
The outputs of two ``black boxes'' are given to the monitor that matches them and decides whether they are consistent.
Aside from minor technical difficulties on organizing co-execution and transforming interfaces,
there is a conceptual problem relating to {\it model abstractness}.
As a rule, a model (tending to be as simple as possible) does not specify the system's behavior accurately,
which makes the output matching awkward.
If the model produces some outputs in some order,
it does not necessarily mean that the system should do it in the same manner~--
the order may differ and some of the ouputs may be omitted.
Before using a model for monitoring one has to specify a priori information on its abstractness and give it to the monitor. 
One of the contributions of this paper is an approach that allows easy adaptation of monitors for models
represented in different abstraction levels.

We consider {\it timed systems}, which react on inputs distributed in time and emit outputs at dedicated time points.
Formally, it means that each event is paired together with a time stamp, identifying when exactly the event happened.
For the {\it discrete-time model}, timed sequences of events can be easily transformed
into ordinary ones by removing time stamps and inserting a special {\it tick} event
in proper positions of the original sequence (as many times as necessary)~\cite{timed-automata}.
Nevertheless, even in that case, it is convenient to suppose that each event is tagged with a time stamp.
Executions of a system and its model are described by {\it timed sequences} over the same alphabet.
Assumptions on the model abstractness allow dynamical generalization of linear sequences into
the {\it partially ordered multisets} consisting of events and {\it time intervals} associated with them.
In general terms, the monitor checks on the fly that an implementation trace
is a linearization of the generalized specification trace (subset of the trace) and
all implementation events satisfy the corresponding time interval constraints.

The rest of the paper is organized as follows.
Section~2 introduces the basic mathematical notions used in the work such as a {\it timed word}, {\it trace} and {\it pomset}.
Section~3 is the main part of the paper, in which the suggested method for timed trace matching is described.
The section formalizes implementation and specification behavior and defines a conformance relation between implementations and specifications.
It also describes a monitoring approach in detail and states its correctness. 
Section~4 outlines our experience in using the proposed approach
for simulation-based verification of industrial hardware designs.
Section~5 is a brief survey of the related work.
Section~6 concludes the paper and discusses some of our future research directions.

\section{Preliminaries}

For the rest of the paper, $\Sigma$ denotes a finite alphabet of {\it events}, while $\mathbb{T}$ denotes a {\it time domain}.
An event might be considered as a set of propositions that identify a situation when the event happens.
A time domain is a totally ordered set with no upper bound,
typically, $\mathbb{N}$ (discrete-time model) or $\mathbb{R}^{\ge 0}$ (continuous-time model).
Sequences of events are called {\it words} (the {\it empty word} is denoted by $\epsilon$). 
Symbols $\Sigma^{*}$ and $\Sigma^{\omega}$ stand for the sets of {\it finite} and {\it infinite} words over $\Sigma$, respectively.
The length of a word $w$ is denoted by $|w|$.
If $u$ and $v$ are two words over the same alphabet and $u$ is finite, then $uv$ denotes their {\it concatenation}.
For $w = uv$ we say that $w$ is a {\it continuation} of $u$ with $v$.

Sometimes, it is useful to structurize events by dividing them into {\it inputs} and {\it outputs}
($\Sigma = I \cup O$) and by introducing a notion of {\it port}~\cite{multiport-fsm}.
Let $P=\{1,2,...,k\}$ and $\mathsf{port}: \Sigma \to P$.
Then, the tuple $\langle \Sigma_1, ..., \Sigma_k \rangle$,
where $\Sigma_p = \mathsf{port}^{-1}(p)$, is called a {\it distributed alphabet}.

\begin{definition}[Timed word -- Alur and Dill~\cite{timed-automata}]
A {\it timed word} $w$ over the alphabet $\Sigma$ and the time domain $\mathbb{T}$
is a sequence $(a_0, t_0)(a_1, t_1) ...$ of {\it timed events} $(a_i,t_i) \in \Sigma \times \mathbb{T}$, satisfying the following constraints:
\begin{enumerate}
\item
for each $i \ge 0$, $t_i < t_{i+1}$ holds ({\it monotonicity});
\item
if the sequence is infinite, for every $t \in \mathbb{T}$ there is some $i \ge 0$, such that $t_i > t$ ({\it progress}).~$\Box$
\end{enumerate}
\end{definition}
Strict monotonicity in the definition above can be weaken to monotonicity
(i.e., it can be required that $t_i \le t_{i+1}$ for all $i \ge 0$)~\cite{timed-automata}.
$(\Sigma \times \mathbb{T})^*$ and $(\Sigma \times \mathbb{T})^\omega$ denote the sets of finite and infinite timed words, respectively.
Note that port partitioning implies an additional constraint on a timed word:
\begin{enumerate}
\setcounter{enumi}{2}
\item
for all $i, j \ge 0$, such that $i \ne j$ and $t_i = t_j$, $\mathsf{port}(e_i) \ne \mathsf{port}(e_j)$ ({\it sequentiality}).
\end{enumerate}

In concurrent systems, the concept of {\it independence} is often in use.
Two events are considered as {\it independent} if they cannot be causally related (i.e., they may happen concurrently).
Events on different ports are usually independent, while those on the same port are dependent.
Concurrent execution can be modeled by partially ordered traces of events,
where incomparable events are supposed to occur in indeterminate order or in parallel~\cite{true-concurrency}.
This intuition underlies two formal models of non-interleaving concurrency:
(1) {\it Mazurkiewicz's trace model}~\cite{trace-theory} and
(2) {\it Pratt's pomset model}~\cite{pomset-model}.
The definitions and their extensions for the timed case are given below.


\begin{definition}[Trace -- Mazurkiewicz~\cite{trace-theory}]
An {\it independence relation} over the alphabet $\Sigma$ is a symmetric and irreflexive relation $\EuScript{I} \subset \Sigma \times \Sigma$.
Given an independence relation $\EuScript{I}$, a pair $\langle \Sigma, \EuScript{I} \rangle$ is called a {\it concurrent alphabet}.
Two words $u$ and $v$ are called {\it Mazurkiewicz equivalent} ($u \equiv_{\EuScript{I}} v$)
iff $u$ can be transformed to $v$ by a finite number of exchanges of adjacent, independent events.
A {\it Mazurkiewicz trace} (or, simply, a {\it trace}) is an equivalence class of words
by the Mazurkiewicz equivalence relation.~$\Box$
\end{definition}

The set of traces over the concurrent alphabet $\langle \Sigma, \EuScript{I} \rangle$ is denoted as $\mathbb{M}(\Sigma, \EuScript{I})$.
Given an independence relation $\EuScript{I}$, the relation
$\EuScript{D} = (\Sigma \times \Sigma) \setminus \EuScript{I}$ is called the {\it dependence relation}.
The {\it length} of a trace $\tau$ (denoted by $|\tau|$) is the length of any of its representatives.
If $w$ is a word, $\lbrack w \rbrack_{\EuScript{I}}$ is the trace that includes $w$ as a representative.
A {\it concatenation} of traces over the same concurrent alphabet $\langle \Sigma, \EuScript{I} \rangle$
is defined by the equality $\lbrack u \rbrack_{\EuScript{I}} \lbrack v \rbrack_{\EuScript{I}} = \lbrack uv \rbrack_{\EuScript{I}}$.
A trace $\sigma$ is called a {\it prefix} of $\tau$ ($\sigma \sqsubseteq \tau$)
iff there exists $\gamma$, such that $\sigma \gamma = \tau$.


\begin{example}[Traces]
Let $\Sigma = \{a, b, c, d\}$ and $\EuScript{I} = \{(a,b), (b,a), (c,d), (d,c)\}$.
Then, some traces are as follows:
$$
\begin{array}{rcl}
\lbrack \epsilon \rbrack_{\EuScript{I}} & = & \{\epsilon\}\\ 
\lbrack ad \rbrack_{\EuScript{I}}       & = & \{ad\}\\
\lbrack ab \rbrack_{\EuScript{I}}       & = & \{ab, ba\}\\
\lbrack abcd \rbrack_{\EuScript{I}}     & = & \{abcd, bacd, abdc, badc\}\\
\end{array}
$$
\end{example}


\begin{definition}[Pomset (partially ordered multiset) -- Pratt~\cite{pomset-model}]
A $\Sigma$-{\it labeled partial order} is a tuple $\langle V, \preceq, \lambda \rangle$,
where $V$ is a finite set of {\it vertices} and $\lambda: V \to \Sigma$ is the {\it labeling function}.
Two $\Sigma$-labeled partial orders are called {\it equivalent} iff they are order- and label-isomorphic
(i.e., they are either equal or differ only in the names of vertices).
A {\it pomset} over the alphabet $\Sigma$ is an isomorphism class of $\Sigma$-labeled partial orders.~$\Box$
\end{definition}

Note that words are equivalent to pomsets with the total order,
while multisets are equivalent to pomsets whose partial order is the equality.
For convenience, we will use a concrete representative (a labeled partial order) to denote the pomset.
There is a number of operations on pomsets, including {\it parallel} and {\it sequential composition}.
Let $\sigma = \langle V, \preceq, \lambda \rangle$ and $\gamma = \langle V', \preceq', \lambda' \rangle$
are pomsets over the same alphabet, such that $V \cap V' = \varnothing$.
Define the pomsets $(\sigma \parallel \gamma)$ and $(\sigma~;~\gamma)$ as follows:
$$
\begin{array}{rcl}
(\sigma \parallel \gamma) & = & \langle V \cup V', \preceq \cup \preceq', \lambda \cup \lambda' \rangle\\
(\sigma~;~\gamma)         & = & \langle V \cup V', \preceq \cup \preceq' \cup (V \times V'), \lambda \cup \lambda'  \rangle\\
\end{array}
$$

\begin{figure}
\centering
\includegraphics[width=0.95\textwidth]{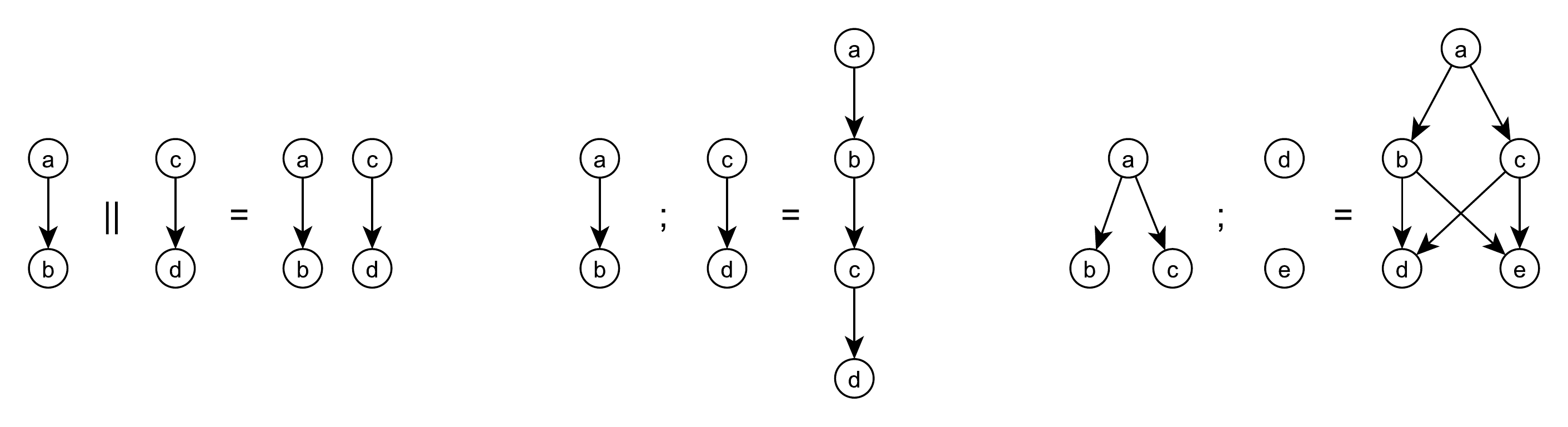}
\caption{Sequential and parallel composition of simple pomsets}
\label{pomset-composition-examples}
\end{figure}

\begin{example}[Pomsets]
Examples of pomsets in the form of {\it Hasse diagrams} (i.e., drawings of the partial order transitive reduction),
may be found in Figure~\ref{pomset-composition-examples}.
\end{example}

A {\it linearization} of a pomset $\langle V, \preceq, \lambda \rangle$ is a total labelled order
$\langle V, \le, \lambda \rangle$, where $\preceq \subseteq \le$.
The set of linearizations of a pomset $\sigma$ is denoted by $\mathsf{lin}(\sigma)$.
A designation $x \bot y$ means that neither $x \preceq y$ nor $y \preceq x$.
We say that $x \in V$ {\it immediately precedes} $y \in V$ and write $x \dot{\prec} y$
iff $x \prec y$ and there is no such $z \in V$ that $x \prec z \prec y$.
A {\it history} of $x \in V$ is the set $\downarrow x = \{y \in V \mid y \preceq x \}$
(for $X \subseteq V$, $\downarrow X = \bigcup_{x \in X}{\downarrow x}$).

It can be shown that each trace can be represented as a pomset.
The opposite is true only for a restricted class of pomsets~\cite{true-concurrency}.
Let $\langle \Sigma, \EuScript{I} \rangle$ be a concurrent alphabet and $\sigma = \langle V, \prec, \lambda \rangle$ be a pomset,
such that 
\begin{itemize}
\item [-]
for each $x \in V$, $\downarrow x$ is a finite set;
\item [-]
for all $x, y \in V$, if $x \bot y$, then $\big{(}\lambda(x), \lambda(y)\big{)} \in \EuScript{I}$;
\item [-]
for all $x, y \in V$, if $x \dot{\prec} y$, then $\big{(}\lambda(x), \lambda(y)\big{)} \in \EuScript{D}$.
\end{itemize}
Then, $\mathsf{lin}(\sigma)$ is a trace over $\langle \Sigma, \EuScript{I} \rangle$ and
$\sigma = \mathsf{pom}(\mathsf{lin}(\sigma))$~\cite{true-concurrency}.
Further, we will represent traces as pomsets satisfying the conditions above.
The same consideration is done in~\cite{leuker-foata,timed-traces}.

\begin{definition}[Timed trace -- Chieu and Hung~\cite{timed-traces}]
A {\it timed trace} over the concurrent alphabet $\langle \Sigma, \EuScript{I} \rangle$ and the time doman $\mathbb{T}$
is a quadruple $\langle V, \preceq, \lambda, \theta \rangle$, where
$\langle V, \preceq, \lambda \rangle$ is a trace over $\langle \Sigma, \EuScript{I} \rangle$ and
$\theta: V \to \mathbb{T}$ is a {\it time function} satisfying the following conditions:
\begin{enumerate}
\item
for all $x, y \in V$, if $x \prec y$, then $\theta(x) < \theta(y)$ ({\it causality});
\item
if the trace is infinite, then for every $t \in \mathbb{T}$ there is a cut $C \subseteq V$,
such that $\mathsf{min}_{x \in C} \{\theta(x)\} \ge t$ ({\it progress}).~$\Box$
\end{enumerate}
\end{definition}

The set of timed traces over the concurrent alphabet $\langle \Sigma, \EuScript{I} \rangle$
and the time domain $\mathbb{T}$ is denoted as $\mathbb{M}_{\theta}(\Sigma, \EuScript{I}, \mathbb{T})$.
Note that timed words are a particular case of timed traces.
Given a non-empty timed trace $\sigma = \langle V, \preceq, \lambda, \theta \rangle$,
$\mathsf{begin}(\sigma) = \mathsf{min}_{x \in V}\{\theta(x)\}$ and
$\mathsf{end}(\sigma) = \mathsf{max}_{x \in V}\{\theta(x)\}$
(if $\sigma$ is infinite, $\mathsf{end}(\sigma) = \infty$);
$\sigma_{[t,t+\Delta{t}]}$ is a sub-trace of $\sigma$
consisting of $x \in V$, such that $\theta(x) \in [t,t+\Delta{t}]$.
Let $\EuScript{TI}(\mathbb{T})$ be the set of time intervals over the time domain $\mathbb{T}$
(i.e., $\EuScript{TI}(\mathbb{T}) = \{ \lbrack t, t+\Delta{t} \rbrack \mid t, t+\Delta{t} \in \mathbb{T}\}$).

\begin{definition}[Time interval trace]
A {\it time interval trace} over the concurrent alphabet $\langle \Sigma, \EuScript{I} \rangle$ and the time doman $\mathbb{T}$
is a quadruple $\sigma = \langle V, \preceq, \lambda, \delta \rangle$, where
$\langle V, \preceq, \lambda \rangle$ is a trace over $\langle \Sigma, \EuScript{I} \rangle$ and
$\delta: V \to \EuScript{TI}(\mathbb{T})$ is a function that associates a time interval to a vertex.
The {\it language} of the time interval trace $\sigma$ is the set
$\EuScript{L}(\sigma) = \{ \langle V, \preceq, \lambda, \theta \rangle \in \mathbb{M}_{\theta}(\Sigma, \EuScript{I}, \mathbb{T})
\mid \forall x \in V~.~\theta(x) \in \delta(x)\}$.~$\Box$
\end{definition}

The set of time interval traces over the concurrent alphabet $\langle \Sigma, \EuScript{I} \rangle$
and the time domain $\mathbb{T}$ is denoted as $\mathbb{M}_{\delta}(\Sigma, \EuScript{I}, \mathbb{T})$.
Futher we will deal with pairs consisting of a timed trace $\sigma$ and a time interval trace $\sigma_\delta$,
such that $\sigma \in \EuScript{L}(\sigma_\delta)$.
Such a pair can be expressed as a quintuple $\langle V, \preceq, \lambda, \theta, \delta \rangle$
and is referred to as an {\it extended time interval trace}.
The set of such traces is denoted as $\mathbb{M}_{\theta \delta}(\Sigma, \EuScript{I}, \mathbb{T})$.


\section{Runtime Verification with Executable Models}

A timed word (more precisely, a timed trace with an empty partial order)
describes a concrete execution of the {\it implementation} under verification,
while an extended time interval trace being more general can be considered as a {\it specification} behavior.
Our goal is to check whether an implementation timed word $w_I \in (\Sigma \times \mathbb{T})^{*(\omega)}$
is conforming to a specification trace $\sigma_S \in \mathbb{M}_{\theta \delta}(\Sigma, \EuScript{I}, \mathbb{T})$.
Note that we are interested in {\it on-the-fly} checking, which means that
a monitor ``lives'' in time and matches two traces in an {\it event-driven} fashion.
{\it Trace acceptance} ({\it verdict}) at a given time point has a three-valued semantics~\cite{bauer-2011}:
(1)~$false$ (an inconsistency has been detected),
(2)~$true$ (the implementation execution has been completed and its trace is conforming to the specification trace) and
(3)~$inconclusive$ (the monitoring is in progress and no inconsistency has been found).

\begin{figure}
\centering
\includegraphics[width=0.95\textwidth]{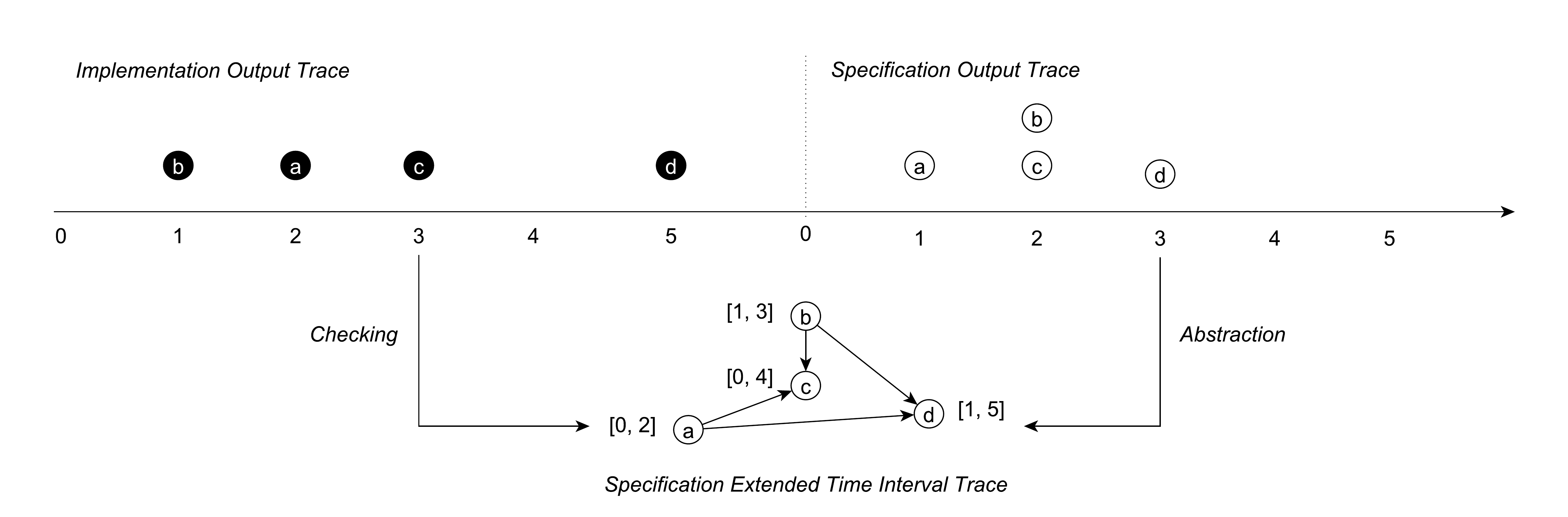}
\caption{Scheme for checking conformance between implementation and specification}
\label{conformance-checking-scheme}
\end{figure}

To make it clear where a specification trace comes from, an additional explanation should be provided.
As it was said in the introduction, a system specification is represented in the {\it executable} form.
Hence, it can be executed and its executions (as ones of the implementation) are represented as timed words.
The straightforward testing of the equality of two timed words is often inadequate
and makes sense only for {\it time-accurate} specifications.
Specifications are usually more abstract than implementations, especially in terms of {\it event ordering} and {\it timing}.
Assumptions on the specification abstractness generalize a concrete timed word to the extended time interval trace softening the conformance checking.
Formally, {\it abstraction} is a map
$\EuScript{A}: (\Sigma \times \mathbb{T})^{*(\omega)} \to \mathbb{M}_{\theta \delta}(\Sigma, \EuScript{I}, \mathbb{T})$,
such that $w$ is conforming to $\EuScript{A}(w)$ for every $w \in (\Sigma \times \mathbb{T})^{*(\omega)}$.
%
%
A specification timed word $w_S$ is mapped into the extended time interval trace $\EuScript{A}(w_S) = \sigma_S$.
Then, it is checked whether an implementation word $w_I$ is conforming to the constructed specification trace $\sigma_S$.
This scheme is illustrated in Figure~\ref{conformance-checking-scheme}.
Technical details can be found in Section~4.

\subsection{Conformance Relation}

The next definition formalizes system executions in terms of timed traces.
It also singles out {\it input} and {\it output sequences} as particular cases of traces corresponding
to stimuli to a system and its reactions, respectively.
System behavior is then abstractly defined as a map of inputs to outputs.

\begin{definition}[Execution trace]
An {\it execution trace} over the concurrent alphabet $\langle \Sigma, \EuScript{I} \rangle$
and the time domain $\mathbb{T}$ is a timed trace with the empty partial order
(i.e., a trace of the kind $\langle V, \varnothing, \lambda, \theta \rangle$).
If $\Sigma = I \cup O$, then execution traces over the alphabet $I$ are called {\it input sequences},
while execution traces over the alphabet $O$ are referred to as {\it output sequences}.~$\Box$
\end{definition}

Note that the empty partial order in execution traces reflects a fact that an implementation is a ``black box'',
and, therefore, the cause-effect relation between its events is unknown.
The sets of input and output sequences are designated by
$\mathbb{I}_{\theta}(\Sigma, \mathbb{T})$ and $\mathbb{O}_{\theta}(\Sigma, \mathbb{T})$, respectively.
Hereinafter, we will use the shortened notations:
$\mathbb{I} = \mathbb{I}_{\theta}(\Sigma, \mathbb{T})$ and
$\mathbb{O} = \mathbb{O}_{\theta}(\Sigma, \mathbb{T})$.


\begin{definition}[Behavior]
{\it Deterministic timed behavior} (or, simply, {\it behavior}) over the alphabet $\Sigma$ and the time domain $\mathbb{T}$
is a (partial) map $\EuScript{B}: \mathbb{I} \times \mathbb{T} \to \mathbb{O}$ satisfying the following constraints:
\begin{itemize}
\item [-]
for every $w \in \mathbb{I}$ and $t \in \mathbb{T}$,
$\mathsf{end}\big{(}\EuScript{B}(w, t)\big{)} \le t$ holds
({\it future uncertainty});
\item [-]
for every $w \in \mathbb{I}$ and $t \in \mathbb{T}$,
$\EuScript{B}(w, t) = \EuScript{B}(w_{[0,t]}, t)$ holds
({\it time directivity});
\item [-]
for every $w \in \mathbb{I}$ and every $t \in \mathbb{T}$,
there exists $w v \in \mathbb{I}$, a continuation of $w$, and $\Delta{t} \ge 0$, such that
$\mathsf{end}\big{(}\EuScript{B}(w v, t + \Delta{t})\big{)} \ge t$
({\it liveness}).~$\Box$
\end{itemize}
\end{definition}

The idea behind the concept is clear.
Behavior describes how an input sequence is transformed into the output sequence taking into account an observation time point.
Usually, when an input sequence is applied, then after a finite number of time units (counting from the last input time)
the output sequence is fully observed and is ready to be checked.
Such post-mortem analysis is not however what we are interested in.
There are two reasons for that:
(1) to ease the analysis, an execution should be terminated as soon as a failure is detected;
(2) storing long sequences in memory is costly.
Providing that a reference model is available, consider how it can be used for checking implementation behavior in runtime.
Let us extend the definition above by allowing a specification to return extended time interval traces over the outputs
(not concrete sequences as it is required).
Denote the set of such traces as $\mathbb{O}_{\theta \delta}$.

Given an output trace $\langle V, \preceq, \lambda, \theta, \delta \rangle \in \mathbb{O}_{\theta \delta}$,
define two functions, $\Delta{t}^{\pm}$, such that
for every $x \in V$, $\delta(x)=\lbrack \theta(x) - \Delta{t}^{-}(x), \theta(x) + \Delta{t}^{+}(x) \rbrack$.
Assume that functions $\Delta{t}^{\pm}$ are bounded
(i.e., there exist constants $\Delta{T}^{\pm} > 0$, such that $|\Delta{t}^{\pm}(x)| \le \Delta{T}^{\pm}$ for all $x \in V$).
Assume also that values $\Delta{t}^{\pm}(x)$ depend only on the event not on the vertex itself
(i.e., $\Delta{t}^{\pm}(x) = \Delta{t}^{\pm}(\lambda(x))$).
Let $\EuScript{I}$ and $\EuScript{S}$ be an {\it implementation} and {\it specification behavior}, respectively.
Given an input sequence $w \in \mathbb{I}$, a time point $t \in \mathbb{T}$,
let us consider implementation and specification outputs:
$\EuScript{I}(w, t) = \langle V_\EuScript{I}, \varnothing, \lambda_\EuScript{I}, \theta_\EuScript{I} \rangle$ and
$\EuScript{S}(w, t) = \langle V_\EuScript{S}, \preceq_\EuScript{S}, \lambda_\EuScript{S}, \theta_\EuScript{S}, \delta_\EuScript{S} \rangle$.
Let us introduce the following notations:
$$
\begin{array}{rcl}
\mathsf{past}^{\Delta{t}}_{\EuScript{I}}(w, t) & = & \{ y \in \EuScript{I}(w, t) \mid \theta_{\EuScript{I}}(y) \le \big{(}t - \Delta{t}^{-}(y)\big{)} \};\\
\mathsf{past}_{\EuScript{I}}(w, t)             & = & \{ y \in \EuScript{I}(w, t) \mid \theta_{\EuScript{I}}(y) \le t \};\\
\mathsf{past}^{\Delta{t}}_{\EuScript{S}}(w, t) & = & \{ x \in \EuScript{S}(w, t) \mid \theta_{\EuScript{S}}(x) \le \big{(}t - \Delta{t}^{+}(x)\big{)} \};\\
\mathsf{past}_{\EuScript{S}}(w, t)             & = & \{ x \in \EuScript{S}(w, t) \mid \theta_{\EuScript{S}}(x) \le t \};\\
\mathsf{match}(x, y)                           & = & \big{(}\lambda_{\EuScript{I}}(y) = \lambda_{\EuScript{S}}(x)\big{)} \wedge \big{(}\theta_{\EuScript{I}}(y) \in \delta_{\EuScript{S}}(x)\big{)}.\\
\end{array}
$$


\begin{definition}[Conformance relation]
The implementation behavior $\EuScript{I}$ is said to be {\it conforming} to the specification behavior $\EuScript{S}$ iff
$\mathsf{dom} \EuScript{I} = \mathsf{dom} \EuScript{S}$ and
for all $w \in \mathsf{dom} \EuScript{S}$ and $t \in \mathbb{T}$,
there is a relation
$\EuScript{M}(w, t) \subseteq \{ (x, y) \in \mathsf{past}_{\EuScript{S}}(w, t) \times \mathsf{past}_{\EuScript{I}}(w, t)
\mid \mathsf{match}(x, y) \}$ (called a {\it matching relation}), such that:
  \begin{enumerate}
  \item
  $\EuScript{M}(w, t)$ is a one-to-one relation;
  \item
  for each $x \in \mathsf{past}^{\Delta{t}}_{\EuScript{S}}(w, t)$, there is $y \in \mathsf{past}_{\EuScript{I}}(w, t)$, such that $(x, y) \in \EuScript{M}(w, t)$;
  \item
  for each $y \in \mathsf{past}^{\Delta{t}}_{\EuScript{I}}(w, t)$, there is $x \in \mathsf{past}_{\EuScript{S}}(w, t)$, such that $(x, y) \in \EuScript{M}(w, t)$;
  \item
  for all $(x, y), (x', y') \in \EuScript{M}(w, t)$, if $x \prec x'$, then $\theta_{\EuScript{I}}(y) \le \theta_{\EuScript{I}}(y')$.
  \end{enumerate}
If for some $w \in \mathbb{I}$ and $t \in \mathbb{T}$ the abovementioned properties are violated,
then $\EuScript{I}$ is said to be {\it not conforming} to $\EuScript{S}$, and $w_{[0, t]}$ is
referred to as a {\it counterexample}.~$\Box$
\end{definition}


\begin{figure}
\centering
\includegraphics[width=0.95\textwidth]{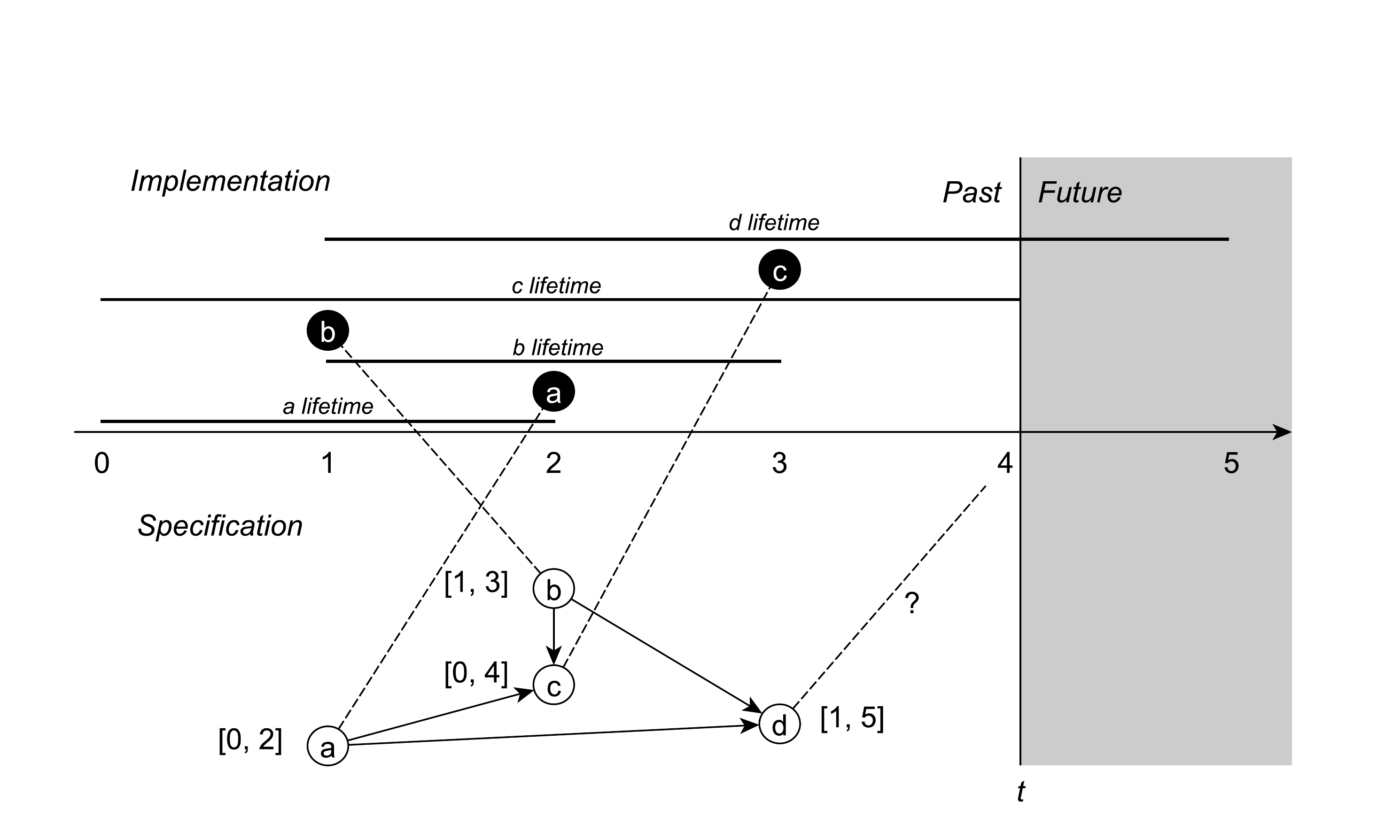}
\caption{Conformance between implementation and specification}
\label{conformance-relation}
\end{figure}

Figure~\ref{conformance-relation} illustrates the conformance relation definition
for a particular input sequence (being unimportant it is not shown in the picture) and observation time ($t = 4$).
The upper part of the figure is a drawing of the implementation outputs (black circles with white labels: $b$, $a$ and $c$).
The lower part depicts the specification outputs (white circles with black labels: $a$, $b$, $c$ and $d$).
Let us denote the trace vertices (i.e., circles themselves) by $y_b$, $y_a$ and $y_c$ (for the implementation) and
$x_a$, $x_b$, $x_c$ and $x_d$ (for the specification).
The implementation vertices are not causally related to each other,
while the specification vertices are partially ordered
(the precedence relation is drawn by arrows:
$x_a \prec x_c$, $x_b \prec x_c$, $x_a \prec x_d$ and $x_b \prec x_d$) and are tagged with time intervals
($\delta(x_a) = [0, 2]$, $\delta(x_b) = [1, 3]$, $\delta(x_c) = [0, 4]$ and $\delta(x_d) = [1, 5]$).
Matchings are depicted by intermittent lines connecting the implementation vertices with the specification ones
($(x_a, y_a)$, $(x_b, y_b)$ and $(x_c, y_c)$).
It is easy to see that this relation fits the matching relation definition:
(1) it is one-to-one relation; (2~\&~3) it includes all events whose lifetime has been exhausted;
(4) is preserves the specification ordering:
\begin{itemize}
\item
$\big{(}x_a \prec x_c\big{)}$ and $\big{(}\theta(y_a) = 2 \le 3 = \theta(y_c)\big{)}$;
\item
$\big{(}x_b \prec x_c\big{)}$ and $\big{(}\theta(y_b) = 1 \le 3 = \theta(y_c)\big{)}$.
\end{itemize}
And, certainly, this relation satisfies the matching condition:
\begin{itemize}
\item
$\big{(}\lambda(x_a) = \lambda(y_a) = a\big{)}$ and $\big{(}\theta(y_a) = 2 \in [0, 2] = \delta(x_a)\big{)}$; 
\item
$\big{(}\lambda(x_b) = \lambda(y_b) = b\big{)}$ and $\big{(}\theta(y_b) = 1 \in [1, 3] = \delta(x_b)\big{)}$; 
\item
$\big{(}\lambda(x_c) = \lambda(y_c) = c\big{)}$ {\tiny ~}and $\big{(}\theta(y_c) = 3 \in [0, 4] = \delta(x_c)\big{)}$. 
\end{itemize}

The next section describes a procedure that automatically and dynamically constructs
a matching relation between implementation and specification outputs.
If it fails to create such a relation, it reports the reason, which can interpreted as a failure type:
a {\it missing} or {\it unexpected} implementation output.


\subsection{On-the-Fly Trace Matching}

A monitor that matches implementation and specification traces and checks their conformance
is co-executed with the implementation and specification and reacts on their outputs.
Formally, the monitor can be expressed as a {\it timed automaton}~\cite{timed-automata} with two types of input ports:
(1) ports for receiving {\it specification outputs} and
(2) ports for receiving {\it implementation outputs}.
When the automaton detects inconsistency between implementation and specification traces,
it goes into a dedicated state informing that the implementation is not conforming to the specification.

A formal description of the on-the-fly trace matcher is given below.
It is represented as a system of {\it guarded actions}.
Each {\it action} is atomic and is executed as soon as the {\it guard} is $true$.
The actions and their guards depend on an external variable $t$ reflecting the current {\it simulation time} and
outputs produced by the specification and implementation in response to the same input sequence ($S$ and $I$, respectively).
The value of $t$ is monotonically increasing in {\it real time}
(simulation time may coincide with real time).
The writing $y \in I{[t]}$ means that at time $t$ the implementation omits an output $y$.
The description is based on two functions:
(1) the {\it primary arbiter} ($arbiter_S$) and (2) the {\it secondary arbiter} ($arbiter_I$),
which are defined as follows:
$$
\begin{array}{rcl}
arbiter_S(X) & = &
\begin{cases}
\mathsf{min}_{\preceq}(X) & \indent\indent\indent\indent\indent\indent \text{if $X \ne \varnothing$,}  \\
\phi                      & \indent\indent\indent\indent\indent\indent \text{otherwise ($\phi \notin \Sigma$);} \\
\end{cases} \\
 & & \\
arbiter_I(y, X) & = & 
\begin{cases}
\mathsf{arg~min}_{x \in X.\mathsf{match}(x, y)}\{ \theta_{\EuScript{S}}(x)\} & \indent \text{if there is $x \in X$, such that $\mathsf{match}(x, y),$} \\
\phi                                                                         & \indent \text{otherwise.} \\
\end{cases} \\
\end{array}
$$
\begin{minipage}{.5\textwidth}
\begin{algorithm}[H]
\caption{$onSpecOutput[x]$, $x \in S{[t]}$}
\begin{algorithmic}
    \Require $true$
    \Ensure $x$
    \State $past_{S} \Leftarrow past_{S} \cup \{x\}$
    \If{$x \in arbiter_S(past_S)$}
        \ForAll{$y \in past_I$ [in ascending of $\theta_{\EuScript{I}}(y)$]}
            \If{$x = arbiter_I(y, \{x\})$}
                \State $past_S \Leftarrow past_S \setminus \{x\}$
                \State $past_I \Leftarrow past_I \setminus \{y\}$
                \State $match~ \Leftarrow match \cup \{ \langle x, y \rangle \}$
                \State trace($\langle x, y \rangle$, \text{``Conforming output''})
                \State {\bf break}
            \EndIf
        \EndFor
    \EndIf
\end{algorithmic}
\end{algorithm}
\end{minipage}
\begin{minipage}{.5\textwidth}
\begin{algorithm}[H]
\caption{$onImplOutput[y]$, $y \in I{[t]}$}
\begin{algorithmic}
    \Require $true$
    \Ensure $y$
    \State $past_{I} \Leftarrow past_{I} \cup \{y\}$
    \State $x \Leftarrow arbiter_I(y, arbiter_S(past_S))$
    \If{$x \ne \phi$}
        \State $past_S \Leftarrow past_S \setminus \{x\}$
        \State $past_I \Leftarrow past_I \setminus \{y\}$
        \State $match  \Leftarrow match \cup \{ \langle x, y \rangle \}$
        \State trace($\langle x, y \rangle$, \text{``Conforming output''})
    \EndIf
    \State
    \State
    \State
    \State
\end{algorithmic}
\end{algorithm}
\end{minipage}
\begin{minipage}{.5\textwidth}
\begin{algorithm}[H]
\caption{$onSpecTimeout[x]$, $x \in past_S$}
\begin{algorithmic}
    \Require $\big{(}\theta_{\EuScript{S}}(x) + \Delta{t}^{+}(x)\big{)} \le t$
    \Ensure $x$
    \State $past_S \Leftarrow past_S \setminus \{x\}$
    \State $verdict \Leftarrow false$`
    \State trace($\langle x, \phi \rangle$, \text{``Missing output''})
    \State ${\bf terminate}$
\end{algorithmic}
\end{algorithm}
\end{minipage}
\begin{minipage}{.5\textwidth}
\begin{algorithm}[H]
\caption{$onImplTimeout[y]$, $y \in past_I$}
\begin{algorithmic}
    \Require $\big{(}\theta_{\EuScript{I}}(y) + \Delta{t}^{-}(y)\big{)} \le t$
    \Ensure $y$
    \State $past_I \Leftarrow past_I \setminus \{y\}$
    \State $verdict \Leftarrow false$
    \State trace($\langle \phi, y \rangle$, \text{``Unexpected output''})
    \State ${\bf terminate}$
\end{algorithmic}
\end{algorithm}
\end{minipage}
\begin{minipage}{.5\textwidth}
\begin{algorithm}[H]
\caption{$onInitialize$}
\begin{algorithmic}
    \Require $t = 0$
    \Ensure $\varnothing$
    \State $past_S \Leftarrow \varnothing$
    \State $past_I \Leftarrow \varnothing$
    \State $match  \Leftarrow \varnothing$
\end{algorithmic}
\end{algorithm}
\end{minipage}
\begin{minipage}{.5\textwidth}
\begin{algorithm}[H]
\caption{$onFinalize$}
\begin{algorithmic}
    \Require $\big{(}{\bf end}(S) + \Delta{T}^{+}\big{)} \le t \wedge
              \big{(}{\bf end}(I) + \Delta{T}^{-}\big{)} \le t$
    \Ensure $\varnothing$
    \State $verdict \Leftarrow true$
    \State ${\bf terminate}$
    \State ~
\end{algorithmic}
\end{algorithm}
\end{minipage}
~\\

Given a time point, the timeout actions ($onSpecTimeout$ and $onImplTimeout$), if they are activated,
are called after the output reception actions ($onSpecOutput$ and $onImplOutput$).
Otherwise, there might be a {\it false negative}.
E.g., when the implementation sends an output $y$ at time $t$
and there is $x \in past_S$, such that $\lambda_{\EuScript{S}}(x) = \lambda_{\EuScript{I}}(y)$ and
$\big{(}\theta_{\EuScript{S}}(x) + \Delta{t}^{+}(x)\big{)} = t$
(thus, $\theta_{\EuScript{I}}(y)$ is a boundary point of $\delta_{\EuScript{S}}(x)$),
calling $onSpecTimeout$ before $onImplOutput$ would lead to the undesirable failure.
If there are two specification outputs $x$ and $x'$, such that $\theta_{\EuScript{S}}(x) = \theta_{\EuScript{S}}(x')$ and $x \prec x'$,
calling $onSpecOutput[x]$ should precede calling $onSpecOutput[x']$. 
The initialization action ($onInitialize$) comes first, while the finalization action ($onFinalize$) is the last action within a time slot.
The order between the timeout actions as the order between the output reception actions is insufficient and may be arbitrary.
The sequence for checking guards and activating actions within a time slot $t$ is as follows:
\begin{enumerate}
\item
initialization ($onInitialize$);
\item
output reception ($onSpecOutput[x]$ and $onImplOutput[y]$, $x \in S[t]$ and $y \in I[t]$);
\item
timeouts ($onSpecTimeout[x]$ and $onImplTimeout[y]$, $x \in past_S$ and $y \in past_I$);
\item
finalization ($onFinalize$).
\end{enumerate}
Note that when we say that some property $\varphi$ holds at time $t$,
we mean that $\varphi$ holds after all of the actions activated at time $t$ have completed.
For a multi-port system, the monitor can be decomposed into a number of loosely connected sub-monitors serving individual ports.
If the specification abstracts away from the inter-port dependencies, the sub-monitors are fully independent and can work in parallel.

\begin{statement}[Monitor correctness]
An input sequence $w$ is a counterexample for $\EuScript{I}$ being conforming to $\EuScript{S}$
iff the monitor terminates with $verdict = false$.~$\Box$
\end{statement}

Rigorously speaking, the termination condition $\big{(}\mathsf{end}(I) + \Delta{T}^{-}\big{)} \le t$
cannot be checked for ``black-box'' implementations
(a monitor is not able to identify whether the implementation is {\it quiescent} or {\it active}).
However, for some types of systems (in particular, systems with {\it covergent behavior})
the condition can be approximated with a checkable one.

\begin{definition}[Convergent behavior]
The behavior $\EuScript{B}: \mathbb{I} \times \mathbb{T} \to \mathbb{O}$ is called {\it convergent}
iff the following conditions are met:
\begin{itemize}
\item [-]
for every finite $w \in \mathbb{I}$, there exists $T(w) \in \mathbb{T}$, called the {\it stabilization time}, such that
for any $t \ge T(w)$, $\EuScript{B}(w, t) = \EuScript{B}\big{(}w, T(w)\big{)}$
($\EuScript{B}(w)$ denotes $\EuScript{B}\big{(}w, T(w)\big{)}$);
\item [-]
for every $t \in \mathbb{T}$, $\EuScript{B}(\epsilon, t) = \epsilon$ holds
(the initial state is {\it quiescent});
\item [-]
for every finite $w, v \in \mathbb{I}$, such that $v \ne \epsilon$ and $t_0 = \mathsf{begin}(v) > T(w)$,
$t \ge t_0$ and $\Delta{t} \in \mathbb{T}$,
$$
\begin{cases}
\EuScript{B}\big{(}w (v + \Delta{t}), t + \Delta{t}\big{)}_{[t_0 + \Delta{t}, t + \Delta{t}]} =
\EuScript{B}(w v, t)_{[t_0, t]} + \Delta{t},& \text{} \\
t_0 \le \mathsf{begin}(\EuScript{B}(w v)_{[T(w), \infty)}),~\text{if $\EuScript{B}(\cdot)_{[\cdot)} \ne \epsilon$;} & \text{} 
\end{cases}
$$
where $w + \Delta{t}$ denotes the sequence constructed from $w$ by adding $\Delta{t}$ to each time stamp of $w$
(quiescent states are {\it stable}).~$\Box$
\end{itemize}
\end{definition}

Assuming that the implementation under verification is convergent,
the termination condition may be expressed as follows:
$$
\Big{(}T(w) \le t\Big{)} \wedge
\Big{(}\big{(}\mathsf{end}(I_{[0,T(w)]}) + \Delta{T}^{-}\big{)} \le t\Big{)}.
$$


\subsection{Specifications with Optional Outputs}

There are systems where operations in some situations terminate other operations,
conflicting with them and of a lower priority.
For example, a write operation can be cancelled by another write operation targeted at the same location and started right after the previous one.
Due to abstractness, a specification is not able to express precisely under what conditions
operations are cancelled and their output is not sent outside.
Taking into account such problems, the definition of the specification behavior should be extended.
Assume there is an unary relation $\Diamond \subseteq V_{\EuScript{S}}$ marking cancellable outputs (the complement of $\Diamond$ is denoted by $\Box$):
if $\Diamond{x}$, then the output is {\it optional} (it might be cancelled,
but the cancellation condition is unknown or inexpressible in specification terms);
if $\Box{x}$, then the output is {\it obligatory} (it cannot be cancelled).
Note that if some action is cancelled, then all dependent actions are cancelled either.

\begin{definition}[Conformance relation for specifications with optional outputs]
The implementation behavior $\EuScript{I}$ is said to be {\it conforming}
to the specification behavior with optional outputs $\EuScript{S}$ iff
${\bf dom} \EuScript{I} = {\bf dom} \EuScript{S}$ and
for all $w \in {\bf dom} \EuScript{S}$ and $t \in \mathbb{T}$,
there is a relation $\EuScript{M}(w, t) \subseteq \{ (x, y) \in
\mathsf{past}_{\EuScript{S}}(w, t) \times \mathsf{past}_{\EuScript{I}}(w, t) \mid \mathsf{match}(x, y) \}$, such that:
  \begin{enumerate}
  \item
  $\EuScript{M}(w, t)$ is a one-to-one relation;
  \item
  for each $x \in \mathsf{past}^{\Delta{t}}_{\EuScript{S}}(w, t)$,
    \begin{itemize}
    \item [-]
    if $\Box{x}$,
    then there is $y \in \mathsf{past}_{\EuScript{I}}(w, t)$,
      such that $(x, y) \in \EuScript{M}(w, t)$;
    \item [-]
    if $\Diamond{x}$, then either
      there is $y \in \mathsf{past}_{\EuScript{I}}(w, t)$,
        such that $(x, y) \in \EuScript{M}(w, t)$, or
      for each $x' \in \mathsf{past}_{\EuScript{S}}(w, t)$,
        if $x \preceq x'$,
        then there is no $y \in \mathsf{past}_{\EuScript{I}}(w, t)$,
        such that $(x', y) \in \EuScript{M}(w, t)$.      
    \end{itemize}
  \item
  for each $y \in \mathsf{past}^{\Delta{t}}_{\EuScript{I}}(w, t)$,
  there is $x \in \mathsf{past}_{\EuScript{S}}(w, t)$, such that $(x, y) \in \EuScript{M}(w, t)$;
  \item
  for all $(x, y), (x', y') \in \EuScript{M}(w, t)$, if $x \prec x'$,
  then $\theta_{\EuScript{I}}(y) \le \theta_{\EuScript{I}}(y')$. $\Box$
  \end{enumerate}
\end{definition}

Checking conformance to specifications with optional outputs can be done with a few modifications of the monitor described above.
In $onSpecTimeout$, it should be checked whether an event $x$ is optional (the action fails only if $x$ is obligatory).
The most difficult part is to track that all events dependent on the cancelled one are also cancelled.
Assume that there is $\Delta{T}_{dep} \in \mathbb{T}$, such that
for all $x, x' \in V_{\EuScript{S}}$, if $|\theta_{\EuScript{S}}(x) - \theta_{\EuScript{S}}(x')| > \Delta{T}_{dep}$, then $x \bot x'$.
To describe the monitor, let us introduce a predicate $cancelled_S(x) = \big{(}\exists x' \in term_S~.~x' \preceq x\big{)}$ and
a modified version of the primary arbiter: $arbiter_S(X) = \mathsf{min}_{\preceq}(X \setminus term_S)$.
~\\
\begin{minipage}{.5\textwidth}
\begin{algorithm}[H]
\caption{$onSpecOutput[x]$, $x \in S{[t]}$}
\begin{algorithmic}
    \Require $true$
    \Ensure $x$
    \State $past_{S} \Leftarrow past_{S} \cup \{x\}$
    \If{$cancelled(x)$}
        \State $term_S \Leftarrow term_S \cup \{x\}$
    \Else
        \State ...
    \EndIf
\end{algorithmic}
\end{algorithm}
\end{minipage}
\begin{minipage}{.5\textwidth}
\begin{algorithm}[H]
\caption{$onSpecTimeout[x]$, $x \in (past_S \setminus term_S)$}
\begin{algorithmic}
    \Require ${(}\theta_{\EuScript{S}}(x) + \Delta{t}^{+}(x){)} \le t$
    \Ensure $x$
    \If{$\Box x$}
        \State ...
    \Else
        \State $term_S \Leftarrow term_S \cup \{x\}$
    \EndIf
    \State
\end{algorithmic}
\end{algorithm}
\end{minipage}
\begin{minipage}{.5\textwidth}
\begin{algorithm}[H]
\caption{$onTermTimeout[x]$, $x \in term_S$}
\begin{algorithmic}
    \Require $\big{(}\theta_{\EuScript{S}}(x) + \Delta{T}_{dep}\big{)} \le t$
    \Ensure $x$
    \State $past_S \Leftarrow past_S \setminus \{x\}$
    \State $term_S \Leftarrow term_S \setminus \{x\}$
\end{algorithmic}
\end{algorithm}
\end{minipage}
\begin{minipage}{.5\textwidth}
\begin{algorithm}[H]
\caption{$onInitialize$}
\begin{algorithmic}
    \Require $t = 0$
    \Ensure $\varnothing$
    \State $term_S \Leftarrow \varnothing$
    \State ...
\end{algorithmic}
\end{algorithm}
\end{minipage}

\section{Tool Support and Experience}

The proposed approach to runtime verification has been implemented in a C++ library named {\it C++TESK Testing ToolKit}~\cite{cpptesk}.
The library provides users with classes and macros for automated development of test system components,
including reference models, monitors (test oracles), stimuli generators, coverage trackers, etc.
C++TESK supports testing and monitoring of both hardware and software systems but has been mainly used for hardware designs
(namely, for simulation-based verification of microprocessor units).
Note that hardware is usually developed in {\it hardware description languages} ({\it HDLs}),
like Verilog and VHDL, and can be executed (simulated) in a special environment, called {\it HDL simulator}.
The C++TESK facilities for developing reference models of hardware designs
(and, consequently, runtime monitors) include means for
sending and receiving data packages,
forking and joining concurrent threads,
modeling time delays and specifying order between data packages.
Some of the primitives (the most important within the scope of the paper) are as follows
(the syntax here differs from the original one, used in the toolkit):
\begin{itemize}
\item
{\tt delay(n)}            --- models a time delay (as an observable outcome, it increments the current time value by {\tt n} time units);
\item
{\tt recv(in):pkg}        --- waits until an input package is received at a given input port ({\tt in});
                              then, returns that package ({\tt pkg});
\item
{\tt send(out, pkg, opt)} --- sends an output package ({\tt pkg}) via a given output port ({\tt out})
                              specifying whether the package is obligatory or optional ({\tt opt}); \\
                              (Note that every time a package is sent outside, it is tagged with time interval
                              $[t - \Delta{t}^{-}, t + \Delta{t}^{+}]$, where $t$ is the sending start time and
                              $\Delta{t}^{\pm}$ are user-defined parameters of the transmission port.)
\item
{\tt depends(pkg1, pkg2)} --- states that an output package ({\tt pkg1}) depends on or causally related to
                              some other package ({\tt pkg2}), input or output. \\
                              (This probably answers the question raised in the beginning of Section~3 of where a specification
                              trace, namely partial ordering of its events, is taken from.)
\end{itemize}

Differences in hardware complexity, verification purposes and amounts of resources
lead to the variety of model types and model abstraction levels.
Abstraction is a well-known way for fighting complexity and facilitating model development. 
Though the verification quality is likely to be lower in case of simpler reference models,
if there is a strict deadline (and it is often so), there is no other way out.
Event ordering and timing are the main subjects for abstraction in hardware designs and other concurrent time-dependent systems.
We use the following classification of the reference models according to the time modeling accuracy:
(1)~{\it untimed models}          (represent only general information on the cause-effect relation of their inputs and outputs,
                                  while the timing is not modelled at all: $\Delta{t}^{\pm} = \infty$),
(2)~{\it time-approximate models} (contain the detailed specification of the event ordering, including some internal arbitration schemes,
                                  but the timing is approximate: $\Delta{t}^{\pm} \le T$, where $T$ has a value of several tens of time units) and
(3)~{\it time-accurate models}    (implement the exact, or almost exact, event ordering and timing: $\Delta{t}^{\pm} \le 1$).

%

The proposed methodology has been used for verification of a number of units of different industrial microprocessors.
Our experience was originally presented in \cite{latw-2011}, and 
since then we have verified a table lookup unit, an l2-cache bank controller and an instruction buffer.
Also, testbenches and monitors for several previously tested components (a north bridge data switch and a memory access unit)
required improvement according to the modifications of the units.
The newest information of the approach application is shown in Table~\ref{experience-table}.
As it can be seen from the table, the methodology supports runtime verification by means of abstract models
(being available at early design stages) and, at the same time, by means of up to time-accurate models
(being available typically at finishing design stages).
Moreover, the approach allows reusing reference models across the hardware development cycle,
which is really important in the industrial settings.

\begin{table}
\centering
\tiny
\begin{tabular}{|p{3.3cm}||p{3.3cm}|p{3.3cm}|p{3.3cm}|}
\hline
\rule{0pt}{12pt}\textbf{Microprocessor Unit} & \textbf{Development Stage} & \multicolumn{2}{|c|}{\textbf{Model Abstraction Level}} \\
\cline{3-4}
 & & \rule{0pt}{12pt}\textbf{From} & \textbf{To} \\
\hline
\hline
\rule{0pt}{10pt}Translation lookaside buffer & Late~/~finishing          & Time-approximate model & Time-accurate model    \\
\hline
\rule{0pt}{10pt}Floating point unit          & Late~/~finishing          & Untimed model          & ---                    \\
\hline
\rule{0pt}{10pt}Non-blocking L2-cache        & Middle~/~late             & Time-approximate model & ---                    \\
\hline
\rule{0pt}{10pt}North bridge data switch     & Middle~/~late~/~finishing & Time-approximate model & Time-accurate model    \\
\hline
\rule{0pt}{10pt}Memory access unit           & Early~/~middle            & Untimed model          & Time-accurate model    \\
\hline
\rule{0pt}{10pt}System interrupt controller  & Early~/~middle            & Untimed model          & Time-approximate model \\
\hline
\rule{0pt}{10pt}Table lookup unit            & Late                      & Time-approximate model & ---                    \\
\hline
\rule{0pt}{10pt}L2-cache bank controller     & Late                      & Time-accurate model    & ---                    \\
\hline
\rule{0pt}{10pt}Instruction buffer           & Late~/~finishing          & Time-accurate model    & ---                    \\
\hline
\end{tabular}
\caption{Experience of the approach application}
\label{experience-table}
\end{table}


The first version of C++TESK supported only accurate reference models
(it was required that a model knows the exact ordering of events on each of the output ports).
Having received feedback from C++TESK users (everyone is welcome to join the community), the toolkit has been modified.
Mostly, it concerns a problem of lack of unit-level specifications even for almost finished hardware designs.
It is impossible to create an accurate model without detailed knowledge of the unit functionality and timing.
Regular interviewing of engineers takes a lot of time and is inconvenient.
Two major solutions of the problem have been proposed besides forcing the developers to write the specifications.
The first solution is to reuse parts of a more complicated system-level model (emulating behavior of the whole microprocessor).
Though such parts are rather abstract (as a rule, system-level models are developed in an untimed manner),
they are really useful for early-stage verification.
The second solution is to develop approximate reference models by means of C++TESK and to refine them if necessary.

\section{Related Work}

There are several works on model-based testing and monitoring that have similarities with our approach.
Some of them are mentioned below.

In \cite{bochmann-2008}, a {\it partial order input/output automaton} ({\it POIOA}),
where each transition is associated with an almost arbitrary ordered set of inputs and outputs,
is used to represent the expected behavior.
The key idea is to obtain two POIOAs (representing behavior of specification and implementation) 
and to check their conformance. There is a way to derive a test suite that guarantees fault detection defined by a POIOA-specific fault model:
{\it missing output faults}, {\it unspecified output faults}, {\it weaker precondition faults}, 
{\it stronger precondition faults} and {\it transfer faults}. 
If the following assumptions are satisfied: an unspecified input is detectable, specified ordering of outputs 
can be observed, response time is bounded, and each specification transition can be modeled as a single implementation transition,
then it is possible to set up conformance between two POIOAs. 
Comforming implementation accepts any input compatible with the specification (and may accept more)
and produces outputs defined by the specification in an order compatible with the specification.
If the POIOAs are not conforming, it is considered as wrong behavior of the implementation according to the fault model.
The main difficulty in the approach, in our opinion, is to represent behavior
of specification and implementation by the proposed formalism.

%

In \cite{passive-testing-2012}, the approach to passive testing based on {\it invariants} is presented.
Invariants are used as a means of representing the most relevant expected properties of the implementation,
which should be exhibited in response to the corresponding test sequences.
Two types of invariants are of usage:
(1) {\it timed consequent invariants} and
(2) {\it timed observational invariants}.
The first type is used to check that an event happens (within certain time bounds) after a given trace of events.
The second type is used to check that a given sequence of events always occur (within certain time bounds) between two given events.
The correctness of the implementation behavior is verified in two steps.
The first step is to check the correctness of the invariants with respect to a given specification.
The second step is to check the correctness of a trace, recorded from the implementation, with respect to the invariants.
We think, that this approach is applicable to monitoring of complex timed systems,
but it is not clear how to maintain the sets of invariants (which might be huge) during the system life cycle.

The approach proposed in \cite{kuliamin-2003} allows usage of implicitly defined {\it asynchronous finite state machines} ({\it AFSMs})
for model-based testing of complex distributed systems.
The implementation behavior is verified only in {\it quiescent states} of the FSM model.
Thus, it is required that there is a predicate identifying such kind of states.
The testing step is done as follows.
First, all outputs are collected and their partial order is determined.
Then, all possible linearizations of the events are enumerated and checked.
If all of them fail (with respect to the specification), then the implementation is not conforming to the specification.
As checking is performed in quiescent states only, the approach is hardly applicable to runtime monitoring
(where there may be arbitrary input sequences, and such states are rarely visited).

\section{Conclusion}

On-the-fly analysis of system behavior is an integral part of dynamic verification of software and hardware systems.
A lot of formalisms have been proposed to express correctness properties for systems of different types,
and a great number of methods have been suggested to check whether system executions are conforming to the specified properties.
None of them is perfect, we think, but all together they cover a vast spectrum of verification and monitoring tasks. 
Among the variety of specification approaches, executable models, written in high-level programming languages, have a significant niche.
First of all, such models are rather universal and allow expressing a broad range of behavioral and structural properties.
Besides, programming languages (especially general-purpose languages, like C and C++) are widely spread in the engineering community.

Our work focuses on using executable models for runtime verification of reactive systems, including, in particular, time-dependent systems.
The problem is not as simple as it looks at first sight.
The naive checking that a system and its model produce the same outputs at the same time is inadequate in the majority of cases.
The model may abstract away from many features implemented in the system under verification such as event ordering and accurate timing
(at least it should be abstracted from the implementation bugs).
We suppose that conformance relations used for runtime verification can be configured in several ways:
(1) by introducing an independency relation over the model events,
(2) by extending time points of the model outputs to time intervals and, finally,
(3) by marking some of the model outputs as being optional.

Basing on this idea, we have developed a method for system monitoring and proved its correctness.
The formalization is based on the theory of traces and partially ordered multisets.
The method has been implemented in C++TESK, an open-source toolkit for hardware modeling, analysis and verification,
and has been successfully used in about 10 projects on simulation-based verification of microprocessor units.
Our future research is aiming at failure diagnostics, which is a deeper analysis of specification and implementation traces being carried out offline.
The goal is to explain what in particular went wrong during the monitoring and give a hint to developers where the bugs are localized. 

\section{Bibliography}

\nocite{*}
\bibliographystyle{eptcs}
\bibliography{generic}

\begin{thebibliography}{10}
\providecommand{\bibitemdeclare}[2]{}
\providecommand{\surnamestart}{}
\providecommand{\surnameend}{}
\providecommand{\urlprefix}{Available at }
\providecommand{\url}[1]{\texttt{#1}}
\providecommand{\href}[2]{\texttt{#2}}
\providecommand{\urlalt}[2]{\href{#1}{#2}}
\providecommand{\doi}[1]{doi:\urlalt{http://dx.doi.org/#1}{#1}}
\providecommand{\bibinfo}[2]{#2}

\bibitemdeclare{}{cpptesk}
\bibitem{cpptesk}
\emph{\bibinfo{title}{C++TESK Homepage}}.
\newblock \urlprefix\url{http://forge.ispras.ru/projects/cpptesk-toolkit/}.

\bibitemdeclare{article}{timed-automata}
\bibitem{timed-automata}
\bibinfo{author}{R.~\surnamestart Alur\surnameend} \& \bibinfo{author}{D.L.
  \surnamestart Dill\surnameend} (\bibinfo{year}{1994}):
  \emph{\bibinfo{title}{A Theory of Timed Automata}}.
\newblock {\sl \bibinfo{journal}{Theoretical Computer Science}}
  \bibinfo{volume}{126}(\bibinfo{number}{2}), pp. \bibinfo{pages}{183--235},
  \doi{10.1016/0304-3975(94)90010-8}.

\bibitemdeclare{article}{passive-testing-2012}
\bibitem{passive-testing-2012}
\bibinfo{author}{C.~\surnamestart Andr\'es\surnameend}, \bibinfo{author}{M.G.
  \surnamestart Merayo\surnameend} \& \bibinfo{author}{M.~\surnamestart
  N{\'{u}}{\~{n}}ez\surnameend} (\bibinfo{year}{2012}):
  \emph{\bibinfo{title}{Formal Passive Testing of Timed Systems: Theory and
  Tools}}.
\newblock {\sl \bibinfo{journal}{Software Testing, Verification \&
  Reliability}} \bibinfo{volume}{22}(\bibinfo{number}{6}), pp.
  \bibinfo{pages}{365--405}, \doi{10.1002/stvr.1464}.

\bibitemdeclare{inproceedings}{barringer-2007}
\bibitem{barringer-2007}
\bibinfo{author}{H.~\surnamestart Barringer\surnameend},
  \bibinfo{author}{D.~\surnamestart Rydeheard\surnameend} \&
  \bibinfo{author}{K.~\surnamestart Havelund\surnameend}
  (\bibinfo{year}{2007}): \emph{\bibinfo{title}{Rule Systems for Run-Time
  Monitoring: From Eagle to RuleR}}.
\newblock In: {\sl \bibinfo{booktitle}{Proceedings of 7$^{th}$ International
  Workshop on Runtime Verification. Revised Selected Papers}}, pp.
  \bibinfo{pages}{111--125}, \doi{10.1007/978-3-540-77395-5\_10}.

\bibitemdeclare{article}{bauer-2011}
\bibitem{bauer-2011}
\bibinfo{author}{A.~\surnamestart Bauer\surnameend},
  \bibinfo{author}{M.~\surnamestart Leucker\surnameend} \&
  \bibinfo{author}{C.~\surnamestart Schallhart\surnameend}
  (\bibinfo{year}{2011}): \emph{\bibinfo{title}{Runtime Verification for LTL
  and TLTL}}.
\newblock {\sl \bibinfo{journal}{ACM Transactions on Software Engineering and
  Methodology}} \bibinfo{volume}{20}(\bibinfo{number}{4}), pp.
  \bibinfo{pages}{14:1--14:64}, \doi{10.1145/2000799.2000800}.

\bibitemdeclare{techreport}{true-concurrency}
\bibitem{true-concurrency}
\bibinfo{author}{B.~\surnamestart Bloom\surnameend} \&
  \bibinfo{author}{M.~\surnamestart Kwiatkowska\surnameend}
  (\bibinfo{year}{1991}): \emph{\bibinfo{title}{Trade-offs in True Concurrency:
  Pomsets and Mazurkiewicz Traces}}.
\newblock \bibinfo{type}{Technical Report TR 91-1223},
  \bibinfo{institution}{Cornell University}.

\bibitemdeclare{inproceedings}{bochmann-2008}
\bibitem{bochmann-2008}
\bibinfo{author}{G.~\surnamestart von Bochmann\surnameend},
  \bibinfo{author}{\surnamestart S.Haar\surnameend},
  \bibinfo{author}{\surnamestart C.Jard\surnameend} \& \bibinfo{author}{G.-V.
  \surnamestart Jourdan\surnameend} (\bibinfo{year}{2008}):
  \emph{\bibinfo{title}{Testing Systems Specified as Partial Order Input/Output
  Automata}}.
\newblock In: {\sl \bibinfo{booktitle}{Proceedings of the 20$^{th}$ IFIP TC
  6/WG 6.1 International Conference on Testing of Software and Communicating
  Systems: 8$^{th}$ International Workshop}}, \bibinfo{series}{TestCom '08 /
  FATES '08}, \bibinfo{publisher}{Springer-Verlag}, \bibinfo{address}{Berlin,
  Heidelberg}, pp. \bibinfo{pages}{169--183},
  \doi{10.1007/978-3-540-68524-1-13}.

\bibitemdeclare{inproceedings}{timed-traces}
\bibitem{timed-traces}
\bibinfo{author}{D.V. \surnamestart Chieu\surnameend} \& \bibinfo{author}{D.V.
  \surnamestart Hung\surnameend} (\bibinfo{year}{2012}):
  \emph{\bibinfo{title}{Timed Traces and Their Applications in Specification
  and Verification of Distributed Real-time Systems}}.
\newblock In: {\sl \bibinfo{booktitle}{Proceedings of the Third Symposium on
  Information and Communication Technology}}, pp. \bibinfo{pages}{31--40},
  \doi{10.1145/2350716.2350723}.

\bibitemdeclare{inproceedings}{latw-2011}
\bibitem{latw-2011}
\bibinfo{author}{M.~\surnamestart Chupilko\surnameend} \&
  \bibinfo{author}{A.~\surnamestart Kamkin\surnameend} (\bibinfo{year}{2011}):
  \emph{\bibinfo{title}{A TLM-Based Approach to Functional Verification of
  Hardware Components at Different Abstraction Levels}}.
\newblock In: {\sl \bibinfo{booktitle}{Proceedings of the 12$^{th}$
  Latin-American Test Workshop}}, pp. \bibinfo{pages}{1--6},
  \doi{10.1109/LATW.2011.5985902}.

\bibitemdeclare{book}{model-checking}
\bibitem{model-checking}
\bibinfo{author}{E.M. \surnamestart Clarke\surnameend},
  \bibinfo{author}{O.~\surnamestart Grumberg\surnameend} \&
  \bibinfo{author}{D.A. \surnamestart Peled\surnameend} (\bibinfo{year}{1999}):
  \emph{\bibinfo{title}{Model Checking}}.
\newblock \bibinfo{publisher}{The MIT Press}.

\bibitemdeclare{article}{spec-testing-2009}
\bibitem{spec-testing-2009}
\bibinfo{author}{R.M. \surnamestart Hierons\surnameend},
  \bibinfo{author}{K.~\surnamestart Bogdanov\surnameend},
  \bibinfo{author}{\surnamestart J.P.Bowen\surnameend},
  \bibinfo{author}{R.~\surnamestart Cleaveland\surnameend},
  \bibinfo{author}{J.~\surnamestart Derrick\surnameend},
  \bibinfo{author}{J.~\surnamestart Dick\surnameend},
  \bibinfo{author}{M.~\surnamestart Gheorghe\surnameend},
  \bibinfo{author}{M.~\surnamestart Harman\surnameend},
  \bibinfo{author}{K.~\surnamestart Kapoor\surnameend},
  \bibinfo{author}{P.~\surnamestart Krause\surnameend},
  \bibinfo{author}{G.~\surnamestart L{\"{u}}ttgen\surnameend},
  \bibinfo{author}{A.J.H. \surnamestart Simons\surnameend},
  \bibinfo{author}{S.~\surnamestart Vilkomir\surnameend}, \bibinfo{author}{M.R.
  \surnamestart Woodward\surnameend} \& \bibinfo{author}{H.~\surnamestart
  Zedan\surnameend} (\bibinfo{year}{2009}): \emph{\bibinfo{title}{Using Formal
  Specifications to Support Testing}}.
\newblock {\sl \bibinfo{journal}{ACM Computing Surveys}}
  \bibinfo{volume}{41}(\bibinfo{number}{2}), pp. \bibinfo{pages}{9:1--9:76},
  \doi{10.1145/1459352.1459354}.

\bibitemdeclare{article}{contract-specs-2007}
\bibitem{contract-specs-2007}
\bibinfo{author}{V.P. \surnamestart Ivannikov\surnameend},
  \bibinfo{author}{A.S. \surnamestart Kamkin\surnameend}, \bibinfo{author}{A.S.
  \surnamestart Kossatchev\surnameend}, \bibinfo{author}{V.V. \surnamestart
  Kuliamin\surnameend} \& \bibinfo{author}{A.K. \surnamestart
  Petrenko\surnameend} (\bibinfo{year}{2007}): \emph{\bibinfo{title}{The Use of
  Contract Specifications for Representing Requirements and for Functional
  Testing of Hardware Models}}.
\newblock {\sl \bibinfo{journal}{Programming and Computer Software}}
  \bibinfo{volume}{33}(\bibinfo{number}{5}), pp. \bibinfo{pages}{272--282},
  \doi{10.1134/S0361768807050039}.

\bibitemdeclare{inproceedings}{kuliamin-2003}
\bibitem{kuliamin-2003}
\bibinfo{author}{V.~\surnamestart Kuliamin\surnameend},
  \bibinfo{author}{A.~\surnamestart Petrenko\surnameend},
  \bibinfo{author}{N.~\surnamestart Pakoulin\surnameend},
  \bibinfo{author}{A.~\surnamestart Kossatchev\surnameend} \&
  \bibinfo{author}{I.~\surnamestart Bourdonov\surnameend}
  (\bibinfo{year}{2003}): \emph{\bibinfo{title}{Integration of Functional and
  Timed Testing of Real-Time and Concurrent Systems}}.
\newblock In \bibinfo{editor}{M.~\surnamestart Broy\surnameend} \&
  \bibinfo{editor}{A.~\surnamestart Zamulin\surnameend}, editors: {\sl
  \bibinfo{booktitle}{Perspectives of System Informatics}}, {\sl
  \bibinfo{series}{Lecture Notes in Computer Science}} \bibinfo{volume}{2890},
  \bibinfo{publisher}{Springer Berlin Heidelberg}, pp.
  \bibinfo{pages}{450--461}, \doi{10.1007/978-3-540-39866-0-45}.

\bibitemdeclare{book}{lam}
\bibitem{lam}
\bibinfo{author}{W.K. \surnamestart Lam\surnameend} (\bibinfo{year}{2005}):
  \emph{\bibinfo{title}{Hardware Design Verification: Simulation and Formal
  Method-Based Approaches}}.
\newblock \bibinfo{publisher}{Prentice Hall}.

\bibitemdeclare{book}{sw-verification}
\bibitem{sw-verification}
\bibinfo{author}{J.~\surnamestart Laski\surnameend} \&
  \bibinfo{author}{W.~\surnamestart Stanley\surnameend} (\bibinfo{year}{2009}):
  \emph{\bibinfo{title}{Software Verification and Analysis: An Integrated,
  Hands-On Approach}}.
\newblock \bibinfo{publisher}{Springer}.

\bibitemdeclare{inproceedings}{leuker-foata}
\bibitem{leuker-foata}
\bibinfo{author}{M.~\surnamestart Leucker\surnameend} (\bibinfo{year}{2000}):
  \emph{\bibinfo{title}{On Model Checking Synchronised Hardware Circuits}}.
\newblock In: {\sl \bibinfo{booktitle}{Proceedings of the 6$^{th}$ Asian
  Computing Science Conference}}, {\sl \bibinfo{series}{Lecture Notes in
  Computer Science}} \bibinfo{volume}{1961}, \bibinfo{publisher}{Springer}, pp.
  \bibinfo{pages}{182--198}, \doi{10.1007/3-540-44464-5\_14}.

\bibitemdeclare{inproceedings}{multiport-fsm}
\bibitem{multiport-fsm}
\bibinfo{author}{G.~\surnamestart Luo\surnameend},
  \bibinfo{author}{R.~\surnamestart Dssouli\surnameend},
  \bibinfo{author}{G.~\surnamestart von Bochmann\surnameend},
  \bibinfo{author}{P.~\surnamestart Venkataram\surnameend} \&
  \bibinfo{author}{A.~\surnamestart Ghedamsi\surnameend}
  (\bibinfo{year}{1993}): \emph{\bibinfo{title}{Generating Synchronizable Test
  Sequences Based On Finite State Machine with Distributed Ports}}.
\newblock In: {\sl \bibinfo{booktitle}{Proceedings of the IFIP Sixth
  International Workshop on Protocol Test Systems}}, pp.
  \bibinfo{pages}{53--68}.

\bibitemdeclare{inproceedings}{trace-theory}
\bibitem{trace-theory}
\bibinfo{author}{A.~\surnamestart Mazurkiewicz\surnameend}
  (\bibinfo{year}{1987}): \emph{\bibinfo{title}{Trace Theory}}.
\newblock In: {\sl \bibinfo{booktitle}{Advances in Petri Nets 1986, Part II on
  Petri Nets: Applications and Relationships to Other Models of Concurrency}},
  \bibinfo{publisher}{Springer-Verlag New York, Inc.}, \bibinfo{address}{New
  York, NY, USA}, pp. \bibinfo{pages}{279--324},
  \doi{10.1007/3-540-17906-2\_30}.

\bibitemdeclare{article}{monitor-based}
\bibitem{monitor-based}
\bibinfo{author}{B.~\surnamestart Patel\surnameend} (\bibinfo{year}{2010}):
  \emph{\bibinfo{title}{A Monitor-Based Approach to Verification}}.
\newblock {\sl \bibinfo{journal}{EE Times}}.

\bibitemdeclare{inproceedings}{pomset-model}
\bibitem{pomset-model}
\bibinfo{author}{V.R. \surnamestart Pratt\surnameend} (\bibinfo{year}{1984}):
  \emph{\bibinfo{title}{The Pomset Model of Parallel Processes: Unifying the
  Temporal and the Spatial}}.
\newblock In: {\sl \bibinfo{booktitle}{Seminar on Concurrency}}, pp.
  \bibinfo{pages}{180--196}, \doi{10.1007/3-540-15670-4\_9}.

\bibitemdeclare{article}{sen-2003}
\bibitem{sen-2003}
\bibinfo{author}{K.~\surnamestart Sen\surnameend} \&
  \bibinfo{author}{G.~\surnamestart Rosu\surnameend} (\bibinfo{year}{2003}):
  \emph{\bibinfo{title}{Generating Optimal Monitors for Extended Regular
  Expressions}}.
\newblock {\sl \bibinfo{journal}{Electronic Notes in Theoretical Computer
  Science}} \bibinfo{volume}{89}(\bibinfo{number}{2}), pp.
  \bibinfo{pages}{162--181}, \doi{10.1016/S1571-0661(04)81051-X}.

\bibitemdeclare{book}{hw-verification}
\bibitem{hw-verification}
\bibinfo{author}{B.~\surnamestart Wile\surnameend},
  \bibinfo{author}{J.~\surnamestart Goss\surnameend} \&
  \bibinfo{author}{W.~\surnamestart Roesner\surnameend} (\bibinfo{year}{2005}):
  \emph{\bibinfo{title}{Comprehensive Functional Verification: The Complete
  Industry Cycle}}.
\newblock \bibinfo{publisher}{Morgan Kaufmann}.

\end{thebibliography}

\end{document}